\documentstyle[11pt]{article}
\def\be{\begin{equation}}
\def\ee{\end{equation}}

\title
{{\hfill\small\tt Theor.\,Math.\,Phys.\,101,\,1200-1204\,(1994)}
\\{\hfill}\\
On a problem posed by Pauli }

\author{B.Z. Moroz,\\
{\small\em Max--Planck--Institut f\"ur Mathematik}\\
{\small\em Gottfried--Claren--Strasse, 26}\\
{\small\em D-53225, Bonn, Germany }\smallskip\\
A.M. Perelomov\\
{\small\em Institute of Theoretical and Experimental Physics}\\
{\small\em B. Cheremushkinskaya, 25}\\{\small\em 117259 Moscow,
Russia}}

\date{}

\begin{document}
\maketitle

\begin{abstract}
\noindent A problem posed in a remark by Pauli is discussed: is it
possible to recover the state vector of a quantum system from the
distribution functions of the physical observables of this system
?
\end{abstract}
\bigskip

{\hfill \small\em Dedicated to the memory of M.K. Polivanov}
\bigskip\medskip

\noindent
Is it possible to recover the state vector of a
quantum-mechanical object from the distribution functions of
measured observables in this state ? This rather naturally posed
problem leads to complicated mathematical problems that are not
currently amenable to solution. We restrict ourselves here to some
simple observations and a precise formulation of the corresponding
mathematical problem.

The possible states of a non-relativistic spinless particle are
described by one-dimensional subspaces of complex Hilbert space
$L^2\,\left( {\bf R}^3 \right)$. Kinematically, one can specify
("measure") the distribution functions of the coordinate operator
${\bf x}$, the momentum operator ${\bf p}$, the angular momentum
operator ${\bf L }= {\bf x} \times {\bf p}$, and the operator of
one of its projections $L_{\nu }$. We denote by $\{ e_{i}^{(\nu )}
\}$ a basis of common eigenfunctions of the operators ${\bf
L}^{2}$, $L_{\nu }$ and expand the state vector $\psi _{j}$
$[j=1,2,\quad \psi _j \in  L^{2}\left( {\bf R}^{3}\right )$] with
respect to this basis:
\[
\psi _{j} = \sum _{i=1}^{\infty }\alpha _{ij}^{(\nu )}e_{i}^{(\nu )}. \]
We assume that
\begin{equation}
\left| \alpha _{i1}^{(\nu )}\right| = \left| \alpha _{i2}^{(\nu )}\right|,
\qquad | \psi _{1}(x)| = |\psi _{2}(x)|,\qquad \left| \hat \psi _{1}(p)
\right| = \left| \hat \psi _{2}(p)\right| \end{equation}
for all $i$, $x$, $p$, where
\[
\hat \psi (p) = \frac1{(2\pi )^{3/2}}\int _{{\bf R}^{3}}\psi (x)\,
\exp (ip^tx)\, dx. \] Does it follow from the relations (1) that
[1]--[4] $\psi _1=\beta \psi _2$, $\beta \in{\bf C}$? Varying $\nu $,
we can divide this problem into the two following problems.

First, it is necessary to describe the set of solutions
\begin{equation}
\left\{ \psi |\psi \in L^{2}({\bf R}^{3}),\quad |\psi (x)| =
W_{1}(x), \quad |\hat \psi (p)| = W_{2}(p)\right\} \end{equation}
for fixed functions $W_{1}$ and  $W_{2}$. This formulation of the
problem is due to Pauli ([5], p.17).

Second, it is necessary to consider self-adjoint operators $A_\nu $,
$1\leq \nu \leq m$, in a finite-dimensional Hilbert space
${\cal H}(\simeq {\bf C}^{n})$. We assume that
\[ A_\nu e_{i\nu }=\lambda _{i\nu }\,e_{i\nu },\qquad 1\leq i\leq n,
\qquad \lambda _{i\nu }\in {\bf C},\]
for bases $\{ e_{i\nu }|1\leq i\leq n\} $ of the space ${\cal H}$.

Let
\[ x=\sum _{i=1}^na_{i\nu }(x)\,e_{i\nu },\qquad a_{i\nu }(x)\in {\bf C},
\quad x \in {\cal H}. \]
How can one describe the set
\[
{\cal A}(b) = \{x\colon |a_{i\nu }(x)| = b_{i\nu },\quad 1\leq i\leq n,
\quad 1\leq \nu \leq m\} \]
for fixed $b=b_{i\nu }$? In particular,  under what conditions on
$\{ A_{\nu }|1\leq \nu \leq m\}$ is the solution of these equations unique
[in ${\bf P}^{n-1}({\bf C})$], i.e., when $x,y \in {\cal A}(b)
\Rightarrow y=\alpha x,\quad \alpha \in {\bf C}$?

We do not know exhaustive answers to these questions, and restrict
ourselves to the following remarks. We refer the interested
readers to the cited literature, in which the physical formulation
of the problem and other results are given.

We begin with the finite-dimensional problem. In this case the following
theorem holds [6].

{\bf Proposition 1.} {\em In the case} $m = 3$, $n\geq 9$, {\em one of the
sets} ${\cal A}(b)$ {\em contains for some $b$ at least two different
solutions}.

{\bf Proof.} Suppose otherwise; then the solution is unique. Therefore,
the mapping
\[ x \mapsto  |a_{i\nu }(x)|,\qquad 1\leq i\leq n,\qquad 1\leq \nu \leq 3,\]
defines a continuous embedding of ${\bf CP}^{n-1}$ in the sphere
$S^{3n-1}$. (Without loss of generality, it can be assumed that
$\sum _{i=1}^{n}|a_{i\nu }(x)|^{2}=1$, for $1\leq \nu \leq 3$).
Since $3n-1>(3/2)(2(n-1)+1)$, there exists a differentiable embedding
of ${\bf CP}^{n-1}$ in $S^{3n-1}$ [7]; therefore, by the well-known theorem
([8], p.390)
\begin{equation}
3n-1>4(n-1)-2\alpha (n-1),\end{equation}
where $\alpha (k)$ is the number of ones in the binary expansion of $k$.
For $n\geq 9$, the inequality (3) does not hold.

In particular, this theorem shows that the distributions of the three
projections of the spin do not uniquely determine the spins state of
the system for sufficiently large (in absolute magnitude) spins.

The second observation is as follows. We choose an odd prime $p$ and
consider the space of functions
\[
X = \{f|f\colon {\bf Z}/p{\bf Z} \rightarrow {\bf C}\}.\]
It is clear that  $X \simeq {\bf C}^{p}$. We take two bases:
\[
\{ \delta _{n}|\,0 \leq n \leq p-1\},\qquad \{ X_{a}| \/ 0 \leq a\leq p-1\},
\]
where
\[
\delta _{n}(m)=\left\{ \begin{array}{ll}0,&n\neq m\cr 1,&n=m,\end{array}
\right.;\qquad X_{a}(m)=\exp \left( \frac{2\pi iam}{p}\right).
\]
Let
\[
\psi _{a}(m) = \exp \left( 2\pi i\frac{am^{2}}{p}\right) ,\qquad
0\leq a\leq p-1.\]
It is clear that $\psi _a\in X$, and therefore
\[
\psi _{a} = \sum _{j=0}^{p-1}b_{aj}\,\delta _{j},\qquad \psi _{a} =
\sum _{j=0}^{p-1}c_{aj}\,X_{j}, \]
with
\[ b_{aj}\in {\bf C},\qquad c_{aj}\in {\bf C},\qquad 0\leq a,j\leq p-1. \]

{\bf Proposition 2.} {\em For $1\leq a \leq p-1$, we have $|c_{aj}| =
1/\sqrt{p}$,  $|b_{aj}| = 1$ for any $j$}.

{\bf Proof.} It is clear that $b_{aj} = \psi _{a}(j)$ and
\[
c_{aj} = \frac1{p} \sum _{m=0}^{p-1}\psi _a(m)\,\overline {X_{j}(m)}. \]
Therefore
\begin{eqnarray*}
|c_{aj}|^{2}\,p^{2}&=&\sum _{0\leq n,m\leq p-1} \psi _a(m)\,
\overline {X_{j}(m)} \overline {\psi_a(n)}\,X_{j}(n) \\
&=& \sum _{0\leq n,m\leq p-1} X_{j}(n-m)\,\exp \left( 2\pi i\,
\frac{a(m^{2}-n^{2})}p\right) .\end{eqnarray*}
Introducing a new variable of summation, we obtain
\begin{eqnarray*}
|c_{aj}|^{2}p^{2}& =&\sum _{0\leq k,n\leq p-1}X_{j}(k)\,X_{j}(-k(2n+k))\\
&=&\sum _{0\leq k\leq p-1}\overline {X_{j}(k^{2})}\,X_{j}(k)
\sum _{0\leq n\leq {p-1}}\overline {X_{j}(2kn)} = p.\end{eqnarray*}

We note the following infinite-dimensional analog of Proposition 2. We set
\[ f_{\alpha }(x) = \exp (i\alpha x^{2}), \]
so that
\[
\hat f_{\alpha }(p) =\frac1{\sqrt {2\alpha i}}\,\exp \left( -\,\frac
{ip^{2}}{4\alpha }\right) .\]
It is clear that the distribution functions $|f_{\alpha }(x)|$,
$|\hat f_{\alpha }(p)|$ do not depend on $\alpha $ for $\alpha \in {\bf R}$.
Unfortunately, however, $f_{\alpha }\not\in L^{2}({\bf R})$. This example was
proposed by  Aharonov [9]. It was noted quite long ago [1], [10]
that if one sets $|\psi (x)|=\rho (x)$, $\psi (x)=\rho (x)\, \exp
(i\varphi (x))$ and defines the function $\psi _1(x)$ by
\[
\psi _1(x)=\rho (x)\, \exp (-i\varphi (-x)), \]
then
\begin{eqnarray*}
\hat \psi _{1}(p) &=& \frac1{\sqrt {2\pi }}\int _{-\infty }^{\infty }
\rho (x)\,\exp (-i\varphi (-x)+ipx)\,dx \\
&=& \frac1{\sqrt {2\pi }}\int _{-\infty }^{\infty }\rho (-x)\,
\exp (-i\varphi (x)-ipx)\,dx
\end{eqnarray*}
and therefore  $\hat \psi _{1}(p) = \overline {\hat \psi (p)}$ if
$\rho (-x) = \rho (x)$.

Thus, choosing $\psi $ in $L^2({\bf R})$, we find
\begin{eqnarray*}
\rho (x) &=& \rho (-x)\quad \mbox{for}\quad x\in {\bf R}\Rightarrow
|\psi (x)| = |\psi _{1}(x)|,\\
|\hat \psi (p)|&=&|\hat \psi _{1}(p)|\quad \mbox{for}\quad x,p\in {\bf R}.
\end{eqnarray*}

Despite the fact that, except for this example, we have not
obtained any results in the case $X=L^2({\bf R})$, one of us
advances the following conjecture:

{\bf Conjecture (A.M. Perelomov)}. {\em Let $\psi ,f \in
L^{2}({\bf R})$. Suppose that $|\psi (x)| = |f(x)|,\quad |\hat
\psi (p)| = |\hat f(p)|$ for almost all $x,p \in {\bf R}$. Then
$f = \alpha \psi $, or $f = \alpha \psi _{1}$ for some $\alpha \in {\bf C}$}.

In the general case, $X = L^{2}({\bf R}^{l})$, the considered example shows
that
\[ |\psi (x)|=|\psi _1(x)|,\qquad |\hat \psi (p)|=|\hat \psi _1(p)| \]
for almost all $x$, $p\in {\bf R}^l$, where
\[
\psi (x)=\psi _{0}(|x|),\quad \psi _{1}(x) = \overline {\psi (x)}, \qquad
\psi _0\in L^2({\bf R}),\qquad |x|\colon= \sqrt {x_{1}^{2}+\cdots +x_{l}^{2}}.
\]
Therefore, in general it does not follow from the relations (1) that
$\psi _{1} = \beta \psi _{2}$ for some $\beta \in {\bf C}$ [1]
(since the angular momentum is zero for spherically symmetric
states).

In conclusion, we describe the subset of the set (2)  that consists of
Gaussian exponentials. We begin with the following simple remark.

{\bf Lemma .} {\em Let $\psi \in L^{2}({\bf R}^{2})$ and $C\in GL(l,
{\bf R})$. Setting $\psi _C(x)\colon =\psi (Cx)$, we have}
\[
\hat \psi _{C}(p) = |\det C|^{-1}\/ \hat \psi ((C^{-1})^{t}p) \]
{\em and}
\[ \Vert \psi _{C}\Vert = |\mbox{det} C|^{-1/2}\Vert \psi \Vert . \]

{\bf Proof.} We have
\begin{eqnarray*}
\hat \psi _{C}(p) &=& (2\pi )^{-{l/2}}\int _{{\bf R}^{l}}\psi (Cx)\,
\exp(ip^{t}x)\,dx \\
&=& |\det C|^{-1}\,(2\pi )^{-{l/2}}\int _{{\bf R}^{l}}\psi (y)\,
\exp (ip^{t}C^{-1}y)\,dy \end{eqnarray*}
and, in addition,
\[
\Vert \psi _{C}\Vert ^{2} = \int _{{\bf R}^{l}}|\psi (Cx)|^{2}\,dx =
|\det C|^{-1}\,\Vert \psi \Vert ^{2}. \]

We now set
\[ \psi (x)=a\,\exp\left( -\frac12\,x^tAx\right) ,\qquad a\in {\bf C},
\qquad A=A^t, \qquad A\in GL(l,{\bf C}) \]
assuming that $\mbox{Re}\,A$ is a positive-definite matrix. Let  $A = A_{1}
+ i\,A_{2}$, $A_{j}\in GL(l,{\bf R})$, $A_j^t=A_j$ for $j=1,2$;
since the matrix $A_{1}$ is positive definite,
\[
C^tAC=I+\mbox{diag}\,(\lambda _1,\ldots ,\lambda _l)
\]
for same $C\in GL(l,{\bf C})$. Therefore
\[ |\hat \psi (p)| = b\,\exp \left( -\frac12\,p^{t}\,B_{1}p\right) ,
\quad B_{1} = \mbox{diag}\,(\mu _{1}^{2},\ldots ,\mu _{l}^{2}), \]
and $0<\mu _{j}\leq 1$ for $1\leq j\leq l$. Thus we set
\[
W_{1}(x) = \pi ^{-l/2}\,\exp \left( -\,\frac12\,x^{t}x\right) ,\qquad
W_{2}(p) = \pi ^{-l/2}b\,\exp \left( -\,\frac12\,p^{t}\,B_{1}p\right) ,\]
choosing
\[ b = \Pi _{j=1}^{l}\,\mu _{j},\qquad B_{1} = \mbox{diag}\,(\mu _{1}^{2},
\ldots ,\mu _{l}^{2}), \]
so that
\[
\psi (x) = W_{1}(x)\exp \left( -\frac12\,ix^{t}\,A_{2}x\right) ,\qquad
\hat \psi (p) =W_{2}(p)\,\exp \left( -\frac12\,ip^{t}\,B_{2}p\right) . \]
The real symmetric matrices $B_1$, $A_2$, $B_2$ satisfy the relations
\[
(I+iA_{2})(B_{1}+iB_{2}) = I, \]
i.e.,
\[ B_{1}-A_{2}B_{2} = I,\qquad A_{2}B_{1}+B_{2}=0. \]
 Thus, it is sufficient to find all (in fact symmetric) solutions of
 the equation $A_{2}^{2} = C$ for $C = (I-B_{1})B_{1}^{-1}$ or
$C = \mbox{diag}\,(\ldots , (1-\mu _{j}^{2})\,\mu _{j}^{-2},\ldots ).$
It follows from this that the general solution for our problem has the form
\[
A_2=\sigma ^t\,D\sigma ,\qquad D\in {\cal L},\qquad \sigma ^t=\sigma ^{-1},
\qquad \sigma B_1=B_1\sigma , \]
where the set
\[
{\cal L} \colon = \{ \mbox{diag}\,(\ldots , \lambda _{j},\ldots )|
\lambda _{j}^{2} = (1-\mu _{j}^{2})\,\mu _{j}^{-2},\qquad 1\leq j\leq l\} \]
contains precisely $2^{l}$ elements. In other words, the set
of solutions decomposes into $2^l$ orbits of the group $G=\{ \sigma |\sigma
\in O(l)$, $\sigma B_1=B_1\sigma \}$.

As a simple example (the idea is due to Kontsevich [11]), we consider
the set of functions
\[\{\psi _\sigma |\psi _\sigma (x)=\psi _0(\sigma x),\qquad \sigma \in O(3)\},
\]
setting
\[
\psi _{0}(x) = a\exp (-\alpha _{1}|x|^{2}-i\,\alpha _{2}(x_{1}^{2}+
x_{2}^{2}-x_{3}^{2})), \]
where
\[ \alpha _{1}>0,\qquad \alpha _{2} \in {\bf R}\setminus \{ 0\},\qquad
x\in {\bf R}^{3}. \]
It is clear that
\[
|\psi _\sigma (x)|=|\psi _0(x)|,\qquad |\hat \psi _\sigma (p)|=
|\hat \psi _0(p)| \]
and  $\psi _{\sigma}\in L^2({\bf R}^3)$ for any $\sigma \in O(3)$;
in particular, it can be seen that the set (2) can be infinite.

One of the authors (B.Z. Moroz) had the possibility of discussing
the questions considered here with colleagues. We take this
opportunity of expressing our deep gratitude to them. We are
especially grateful to Y.Aharonov, A.Connes, M.L. Gromov, F.
Hirzebruch and M.L. Kontsevich, who made some important remarks.

\end{document}